\documentclass[conference]{IEEEtran}

\usepackage{amsmath}
\usepackage{amsfonts}
\usepackage{amssymb}
\usepackage{algorithm}
\usepackage[noend]{algpseudocode}
\usepackage[justification=centering]{caption}
\usepackage[skip=10pt,font=small]{caption}
\usepackage{wrapfig}
\usepackage{graphicx}
\usepackage{float}
\usepackage{xcolor}
\usepackage{soul}
\usepackage{cite}
\usepackage{cancel}
\usepackage{comment}
\usepackage{graphicx}
\usepackage{xcolor}

\usepackage[normalem]{ulem}

\def\BibTeX{{\rm B\kern-.05em{\sc i\kern-.025em b}\kern-.08em
    T\kern-.1667em\lower.7ex\hbox{E}\kern-.125emX}}

\begin{document}

\title{Intelligent O-RAN Traffic Steering for URLLC Through Deep Reinforcement Learning}

\author{
    \IEEEauthorblockN{Ibrahim Tamim, Sam Aleyadeh, and Abdallah Shami}
    \IEEEauthorblockA{Electrical and Computer Engineering Department, Western University, London ON, Canada
    \\ \{itamim, saleyade, abdallah.shami\}@uwo.ca}
}

\maketitle


\begin{abstract}
The goal of Next-Generation Networks is to improve upon the current networking paradigm, especially in providing higher data rates, near-real-time latencies, and near-perfect quality of service. However, existing radio access network (RAN) architectures lack sufficient flexibility and intelligence to meet those demands. Open RAN (O-RAN) is a promising paradigm for building a virtualized and intelligent RAN architecture. This paper presents a Machine Learning (ML)-based Traffic Steering (TS) scheme to predict network congestion and then proactively steer O-RAN traffic to avoid it and reduce the expected queuing delay. To achieve this, we propose an optimized setup focusing on safeguarding both latency and reliability to serve URLLC applications. The proposed solution consists of a two-tiered ML strategy based on Naive Bayes Classifier and deep $Q$-learning. Our solution is evaluated against traditional reactive TS approaches that are offered as xApps in O-RAN and shows an average of 15.81 percent decrease in queuing delay across all deployed SFCs. 

\end{abstract}
\bigskip
\section{Introduction}

Next-Generation Networks (NGNs) are expected to fill the current performance gap between service demands and networking capabilities. Three main services are currently the target of focus in 5G, namely, Enhanced Mobile Broadband (eMBB), massive Machine-Type Communications (mMTC), and Ultra Reliability Low-Latency Communications (URLLC), supporting traffic requiring extremely high reliability (\textit{i.e.}, 99.999\%) while restricted to very low latency (\textit{i.e.}, less than 1 $ms$). Since existing RAN solutions are inflexible, closed, and aggregated, enabling the above services with static approaches is extremely challenging. In recent advances in the Centralized RAN (CRAN) and virtualized RAN (vRAN), attempts are being made to address these issues. However, the lack of open interfaces and non-proprietary hardware and software is hindering any significant breakthroughs. Therefore, Open RAN (O-RAN) has been recently proposed to address these challenges, providing an abundance of flexibility and openness needed through virtualized, disaggregated, and open architecture \cite{polese2022understanding}.

The key objective of O-RAN is to support 5G systems and improve traditional RAN performance by leveraging its elements' virtualization and adopting Machine Learning (ML)/Artificial Intelligence (AI) techniques into its core structure \cite{bonati2021intelligence}. Two novel modules, the near-Real-Time RAN Intelligent Controller (near-RT RIC), and non-RT RIC were developed and tasked with centralized network abstraction to reduce further cost, network complexity, and human interaction \cite{lee2020hosting}. In addition, these modules enable ML-based algorithms to be introduced to any layer of the RAN. 
Another key challenge in 5G is that mobile networks are becoming increasingly complicated as network capacity and traffic is increasing exponentially. With that said, traditional reactive traffic management lacks fundamental attributes barring the serving of stringent demands such as those of URLLC, especially when existing in heterogeneous traffic streams \cite{8428419}. Network virtualization allows for more control over the routing. This gave rise to the field of Traffic Steering (TS), to build virtualized networking architecture to allocate resources to various wireless services with different requirements \cite{dryjanski2021toward}, creating a continuously dynamic and customized solution to match the incoming demands. 

The combination of highly diverse demands and the sheer amount of services offered require special techniques, especially when placed in a dynamic virtualized environment. Therefore, the existing TS concept was retooled to optimally allocate resources to the various wireless services, creating a continuously dynamic customized solution to match the incoming demands. 
\begin{figure}[t]
    \centering
    \setlength{\belowcaptionskip}{-20pt} 
    \includegraphics[width=\columnwidth]{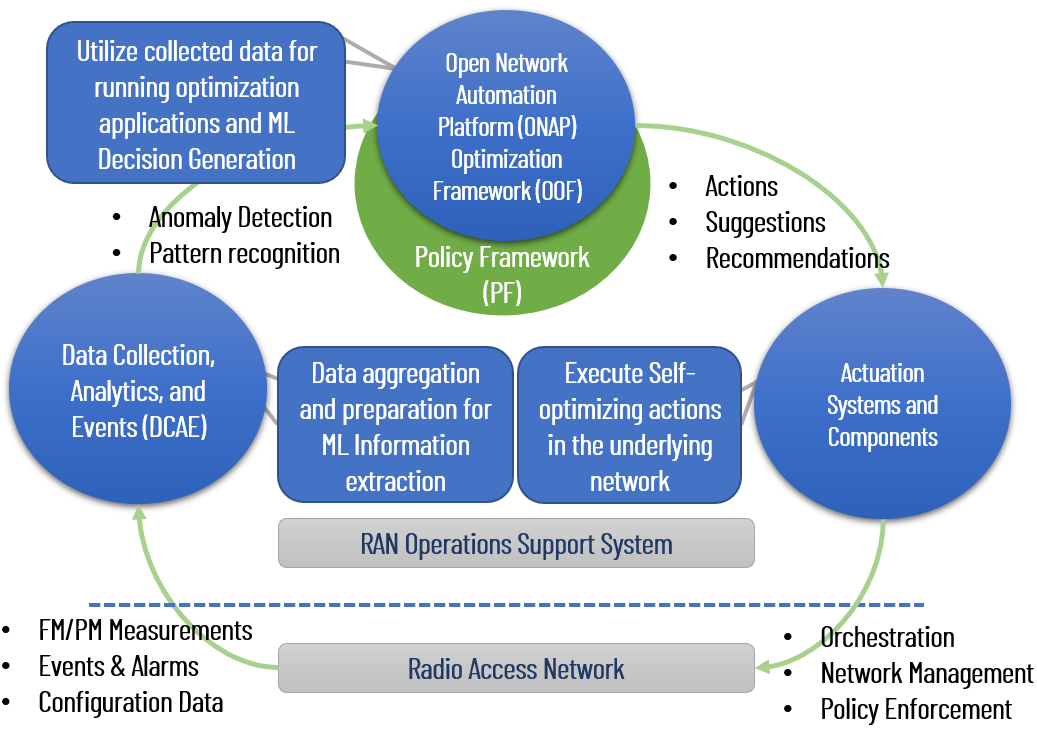}
    \caption{ML enabling cycle in O-RAN}
    \label{fig:ml_in_o-ran}
\end{figure}
TS brings significant benefits to many virtualization-based networking paradigms such as Software-Defined Networking (SDN), and significantly aids in load balancing and resource allocation in Network Function Virtualization (NFV) environments leading to massive energy and cost cuts, thus proving it essential to the O-RAN architecture as well. 
To adopt TS techniques in O-RAN, we can leverage O-RAN's intelligent nature to improve existing TS techniques further to better serve the NGN services mentioned above.
\begin{figure*}[t]
  \includegraphics[scale=0.43]{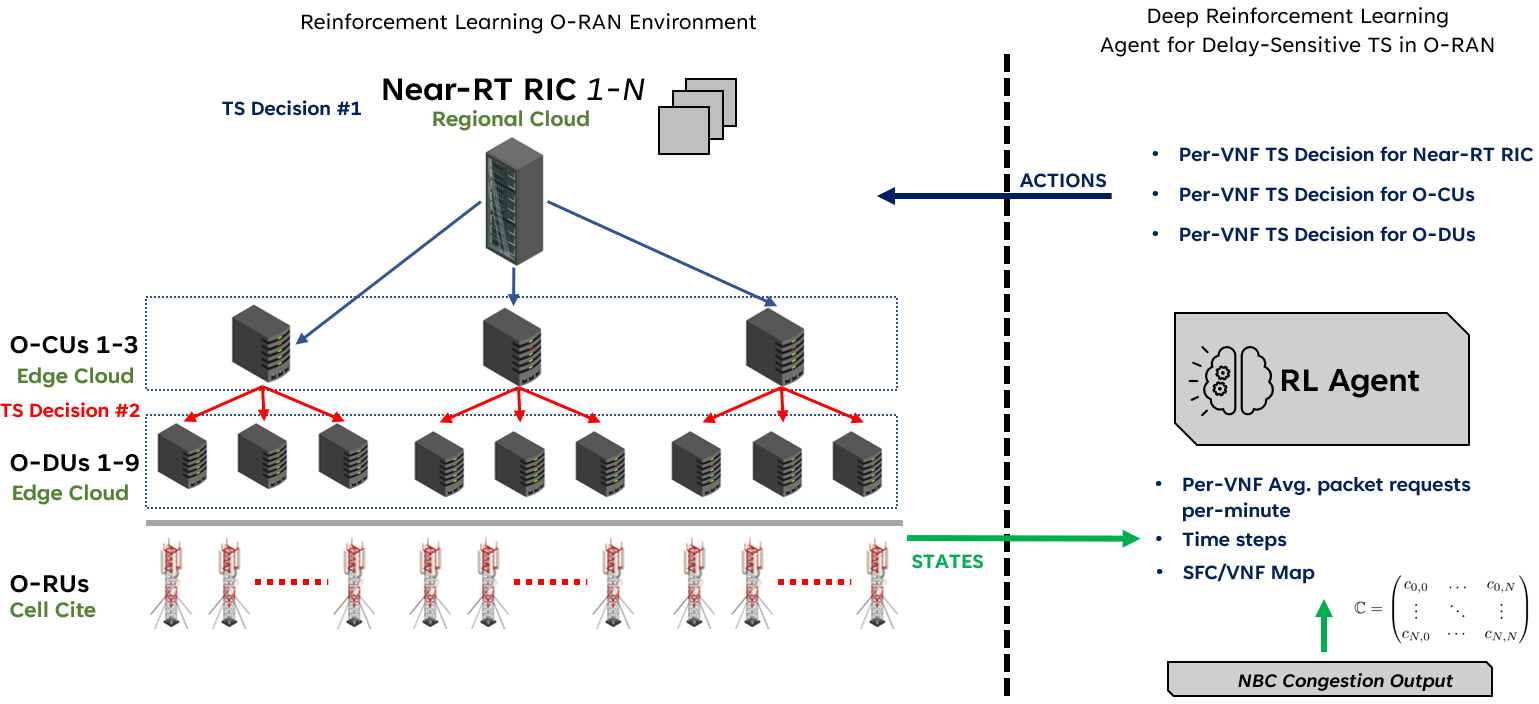}
  \centering
  \setlength{\belowcaptionskip}{-18pt} 
  \caption{Traffic steering RL system overview}
    \label{fig:systemoverview}
\end{figure*}
In this work, we propose a two-tier ML-assisted strategy based on Naive Bayes Classifier (NBC) and deep $Q$-learning. The strategy aims to first predict congestion on every O-RAN Virtualized Network Function (VNF) in a URLLC-specific O-RAN deployment. The strategy then uses these predictions as inputs to a deep $Q$-learning agent that incorporates the time of day to optimally issue TS decisions on any Service Function Chain (SFC) projected to face congestion. The TS decision complies with O-RAN's operational and functional constraints and ensures that the queuing delay across the O-RAN is reduced. In addition, the O-RAN Reinforcement Learning (RL) environment ensures that all the reliability constraints of O-RAN for a URLLC deployment are met to make this solution suitable for critical applications. Finally, the solution follows the ML cycle defined for O-RAN and benefits from O-RAN data collection modules as shown in Fig. \ref{fig:ml_in_o-ran}.

The remainder of this paper is organized as follows. In Section II, we review the state-of-the-art literature. The motivation and problem definition are outlined in Section III. The ML-based system's mathematical model is discussed in-depth in Sections IV and V. Section VI presents and discusses our simulations and the experimental results. Section VII concludes the paper.

\section{Literature Review}
There is a considerable amount of ML-based resource allocation solutions focused on virtualized networking in the literature \cite{9695955}. However, TS-related ML solutions are still scarce. Traffic engineering in virtualized networks requires the consideration of network function placement and traffic routing, which is NP-hard \cite{yu2017qos}. Given the high complexity of the problem, ML solutions became the next natural solution. For example Adamczyk \emph{et al.} \cite{adamczyk2021reinforcement} use  a neural network with the SARSA algorithm for TS within a RAN environment. The solution allocates available resources among the users according to each base station load.  They  define  different  user  profiles  according  to  their data rate demands.

In \cite{pham2021traffic}, Pham formulate their TS solution as an optimization problem in the form of a Mixed Integer Linear Programming (MILP) model. The model captures the essential aspects of TS in NFV, including the optimal VNF placements and routings under constraints on delay-guaranteed SFC to maximize energy efficiency. The authors develop their algorithm based on soft actor-critic that provides an efficient traffic engineering solution for large-scale NFV systems.

Most of the above solutions rely on approximations, especially for large-scale networks; aside from such approaches, heuristic-based approaches are being pursued but focus on limited application-specific scenarios. Tajiki \emph{et al.} also tackle the TS problem from both the routing and initial placement using a heuristic approach focused on the energy consumption of SFCs \cite{8480442}.

Kato \emph{et al.}\cite{kato2016deep} present a deep belief network-based deep learning solution to improve heterogeneous network TS. Mao \emph{et al.} \cite{mao2017routing} propose a deep learning-based routing targeting high-speed core networks. When facing exponentially increasing traffic, the authors further propose a tensor-based deep learning architecture to achieve fine-grained network traffic control \cite{mao2017tensor}.

Koichi \emph{et al.} \cite{adachi2016q} propose a deep learning-based TS mechanism between 3GPP LTE and Wi-Fi networks. The main objective of this mechanism is to provide improved network quality of service without explicit information input from the network side in a dynamically changing HetNet environment and by relying on the limited User Entity (UE) information to estimate the traffic condition of its network and learn about the traffic load and the connection quality. The ML agent then takes actions to reduce outages with reduced network switch usage.

The above solutions have proved their success within their individual scopes but shared common drawbacks, such as the lack of scalability, which is more pronounced in traditional optimization-based techniques. Portability issues are shown in the heuristic-based solutions due to the specificity of the solutions pursued. The combination of these aspects makes it clear that such TS approaches cannot be ported to an URLLC-specific O-RAN environment successfully. Our solution outlined in the following sections aims to leverage AI techniques to address these drawbacks while being tailored for deployment as an rApp (third-party application hosted on the non-RT RIC) or xApp (third-party applications hosted on the near-RT RIC). We achieve this by designing the solution following O-RAN's ML-enabling pipeline and adhering to O-RAN operational and functional constraints. 
\section{Motivation and Problem Definition}
Mission-critical 5G applications such Automatic Vehicle Locator (AVL) for task-sensitive vehicles, and latency/reliability-based applications such as Augmented Reality (AR)-assisted surgery, Vehicle-to-Everything (V2X) communications, live sports events streaming, and online gaming are the leading areas that benefited from the 5G revolution and are at the center of further development in Beyond 5G (B5G) and 6G networks under the use case of URLLC. However, network traffic and user connectivity are expected to grow rapidly in the upcoming years as the number of mobile 5G subscriptions worldwide is expected to reach 4.8B by 2026 and pose further challenges for these networks \cite{statista}.

The O-RAN is one of those components that is being revolutionized in 5G, B5G, and 6G by bringing the power of intelligence and openness to RANs. The continuing development of the O-RAN allows Network Service Providers (NSPs) to utilize advanced ML-based solutions in the RAN level of 5G when they were only possible in the 5G core.

 As the O-RAN architecture enables the use of AI and ML across all its layers, it is currently considered one of the best architectures of choice for the deployment of RAN-specific intelligent solutions. However, O-RAN has its own functional and operational constraints that must be considered at all times for a correct and efficient deployment. When these constraints are met, O-RAN  allows the complete virtualization of all its components \cite{deployment}. In most O-RAN deployment scenarios defined by the O-RAN Alliance, the near-RT RIC, O-RAN Central Unit (O-CU), and O-RAN Distributed Unit (O-DU) are the most common components to always be hosted as VNFs.
 
 Through  virtualization, the O-RAN components can be hosted on Commercial Off-The-Shelf-Servers (COTS) in edge, cloud, and core server locations. The details of all suggested deployment scenarios and the use cases that they are best suited for can be found in \cite{deployment}. In this virtualized environment, the end user requests or specific network services are delivered through the chaining of those VNFs to create O-RAN SFCs. Through this work, we consider an SFC to be the chaining of the monitoring and management VNF, namely the near-RT RIC, and the traffic and packet handling VNFs, namely the O-CU and O-DU type VNFs. 

For optimal operation of the O-RAN, the placement of those VNFs must be optimized to meet a certain objective or optimization goal. NSPs can use several approaches to achieve this initial placement depending on the use case at hand. In \cite{globecomtamim} \cite{iwxcmc}, we present optimization-based solutions to optimize that placement for latency and reliability. 

These placements can greatly improve the operation of the VNFs individually and the SFCs collectively. However, when those SFCs face continuously changing traffic demands and dynamic user behaviours, the O-RAN can suffer from significant performance degradation due to the congestion of individual VNFs. This leads to latency violations that mainly occur due to increasing queuing delays at each VNF. Queuing delay is the time a packet waits in a VNF queue until it can be processed. A VNF can become congested due to the exhaustion of its CPU, memory, or bandwidth resources. This hinders its ability to process packets in a timely manner, leading to violations in latency constraints. These violations are considered extremely serious in URLLC and can result in negative outcomes ranging from financial damages to even endangering human lives. 

In this work, we tackle the weaknesses of the above-mentioned approaches and design a comprehensive end-to-end solution as an xApp deployable directly in the O-RAN architecture. We propose a two-tier algorithm that decouples traffic congestion prediction from the task of TS. We harness the advantages of supervised ML in analyzing network traffic to predict the congestion on each VNF. We then develop a Deep Reinforcement Learning (DRL) agent to steer the traffic proactively to avoid congested VNFs. 
Our DRL agent is also tasked with optimizing the new SFCs for latency and preserving reliability. The agent considers VNFs that meet O-RAN's queuing time thresholds and ensure no reliability constraints are violated by carrying the steering through servers with Mean Time to Fail (MTTF) values following the URLLC thresholds. We achieve our objectives by designing a per-VNF traffic congestion NBC capable of predicting if a VNF will become congested. The NBC's output is inputted to the O-RAN RL environment to be processed by the second tier of the algorithm, a deep $Q$-leaning agent designed to ensure each SFC in the O-RAN operates with reduced queuing delay and follows the reliability constraints. The agent achieves this by issuing TS decisions proactively, avoiding congested VNFs. Fig. \ref{fig:systemoverview} shows an overview of the entire system model with the O-RAN deployment. 

To evaluate our end-to-end solution, we develop a discrete-time event simulator for URLLC O-RAN traffic and test both the NBC and the $Q$-learning agent under dynamic network conditions. The following sections discuss the design of the NBC and the DRL agent, followed by our results and discussions. 
\section{NBC for Congestion Prediction}
\label{nbcsection}
As shown in Fig.\ref{fig:ml_in_o-ran}, O-RAN's Data Collection, Analytics, and Events (DCAE) component monitors and collects key parameters from  O-RAN's networking components, such as the O-CU and O-DU. The DCAE is responsible for data aggregation and preparation for ML information extraction or model training.

In our work, we design an O-RAN traffic simulator and generate URLLC UE traffic that is collected by the DCAE. The collected data is prepared by the DCAE for training. The simulated traffic is composed of three traffic types, namely, augmented/virtual reality, autonomous/guided vehicles, and automated industry. For each, the error rates thresholds are between $10^{-3}$ and $10^{-5}$, greater than or equal to $10^{-3}$, and between $10^{-5}$ and $10^{-9}$, respectively \cite{TRAFFIC}. We also define the maximum tolerable latency for each following URLLC requirements as between $5ms$ and $10ms$ for both augmented/virtual reality and autonomous/guided vehicles, and $1ms$ for automated industry traffic \cite{TRAFFIC}. 

One of the challenges of RANs in general, is the cost and availability of processing units in terms of CPU and memory in edge cloud locations and on resource-scarce components such as O-CUs and O-DUs. This fact becomes more challenging when additional data-gathering modules and ML training and inference hosts are introduced in the O-RAN. To keep our solution computationally feasible and cost efficient, the prediction must be carried out using a lightweight ML algorithm. The data is pre-processed, and the average of packets processed per-minute is calculated from the traffic to be used as an indicator for congested VNFs. The defined one-minute window reduces the load on the training modules. 

The choice of NBC is motivated by the fact that the DCAE for our solution calculates the average packets per minute and the average processing latency for each VNF. Those two variables are independent, and in our simulator, the average packets per minute follow a normal distribution (in most of the timesteps during the day, the loads are normal, spiking during rush hours and becoming lighter during the night). NBCs are optimal for such a setup and are lightweight. This allows us to host the training module in the near-RT RIC regardless if the near-RT RIC was hosted on the regional or edge clouds. The simulations and results for congestion predictions are discussed in Section \ref{results}. 

The key observation in NBCs is that both the features we collect from the DCAE are independent. This means that we are able to assume class-specific covariance matrices, and these matrices are diagonal. Using Gaussian NBC, we assume that the class-conditional densities are normally distributed.  

In Eq. \ref{eq:nbc1}, the algorithm is provided with K=2 input variables $x$ and corresponding target variable $t$. $\mu$ is the class-specific mean vector, and $\sigma$ is the class-specific covariance matrix. With Equation \ref{eq:nbc1} and using Bayes’ theorem, we calculate the class posterior in Equation \ref{eq:vnf_availability}, then $x$ is classified into a class in Equation \ref{eq:vnf_availability 2}. The two classes are \textit{congested} and \textit{not congested}.
\begin{equation} \label{eq:nbc1}
P(x|t=c,{{\mu }_{c}},{{\Sigma }_{c}})=K(x|{{\mu }_{c}},{{\Sigma }_{c}})
\end{equation}

The training is done on 1-minute intervals for packet averages and average latencies. We train the NBC for each VNF, and its lightweight allows for a training host to be deployed with each near-RT RIC, O-CU, and O-DU VNF types with minimal overhead. The resulting output during O-RAN's operation is a congestion classification based on the likelihood of the VNF to become congested. This output is produced for all VNFs in the O-RAN and is then passed to the RL environment for TS and optimization. The following section covers the RL environment and the $Q$-learning agent, followed by the detailed results for both tiers. 
\begin{equation} \label{eq:vnf_availability}
\overbrace{P(t=c|x,{{\mu }_{c}},{{\Sigma }_{c}})}^{Class\,\,\,posterior}=\frac{\overbrace{P(x|t=c,{{\mu }_{c}},{{\Sigma }_{c}}}^{Class-conditional\,density}\,\overbrace{P(t=c)}^{Class\,\,prior}}{\sum\limits_{k=1}^{K}{P(t=k|x,{{\mu }_{k}},{{\Sigma }_{k}})P(t=k)}}
\end{equation}
\begin{equation} \label{eq:vnf_availability 2}
	\overset{\wedge }{\mathop{h(x)}}\,=\underset{c}{\mathop{\arg \max }}\,P(t=c|x,{{\mu }_{c}},{{\Sigma }_{c}})
\end{equation}
\section{Deep $Q$-Learning for URLLC TS}
The core of our solution is the ability to intelligently (using congestion prediction, time of day, and type and behaviour of the VNF) reduce the queuing delay that URLLC packets experience by proactively steering the traffic to the optimal available SFC at any point in time. Optimization, heuristic, and traditional ML approaches suffer many weaknesses when it comes to changing traffic conditions, VNF infrastructure changes, and meeting critical functional and operational constraints as we mentioned above. Specifically in O-RAN, state-of-the-art research has shown the power RL can bring when dealing with time-critical dynamic deployments. However, to design and implement a successful and efficient RL agent, a functional O-RAN RL environment must be designed for rapid simulation. In this section, we cover the design of a URLLC use-case-specific O-RAN RL environment, and the design of a Deep $Q$-Network agent with the goal of achieving the aforementioned objectives. 
\subsection{O-RAN RL Environment and its state and action spaces}
To model and simulate the O-RAN, the RL environment should model both an infrastructure for COTS and NFV. We first define a server infrastructure adjacency matrix $\mathbb{S}$ (Eq. \ref{eq:S}). In $\mathbb{S}$, an ${{s}_{i,j}}$ is 1 if a valid connection between servers $i$ and $j$ exists and 0 otherwise.
\begin{equation} \label{eq:S}
\mathbb{S}=\left( \begin{matrix}
   {{s}_{0,0}} & \ldots  & {{s}_{0,N}}  \\
   \vdots  & \ddots  & \vdots   \\
   {{s}_{N,0}} & \cdots  & {{s}_{N,N}}  \\
\end{matrix} \right)
\end{equation}
The latencies of active links are given in $\mathbb{L}$. $\mathbb{S}$ and $\mathbb{L}$ are the key matrices in determining the initial placement decisions of all VNFs and later determining if a TS decision is valid by checking the latency threshold of each URLLC SFC versus the latencies in $\mathbb{L}$.  
\begin{equation} \label{eq:latency}
\mathbb{L}=\left( \begin{matrix}
   {{l}_{0,0}} & \ldots  & {{l}_{0,N}}  \\
   \vdots  & \ddots  & \vdots   \\
   {{l}_{N,0}} & \cdots  & {{l}_{N,N}}  \\
\end{matrix} \right)
\end{equation}
The NBC classifier outputs a congestion matrix $\mathbb{C}$ (Eq. \ref{eq:cMat}) where ${{c}_{i,j}}$ is 1 if a VNF $j$ is congested. $\mathbb{C}$ is defined as an adjacency matrix as one VNF can belong to several SFCs. In $\mathbb{C}$, if the same VNF $j$ is equal to 1 in multiple indices then this VNF is considered a shared VNF. This approach allows our solution to give higher rewards when alleviating congestion from a shared VNF and more severe penalties for unnecessary TS decisions on these VNFs. 
\begin{equation} \label{eq:cMat}
\mathbb{C}=\left( \begin{matrix}
   {{c}_{0,0}} & \ldots  & {{c}_{0,N}}  \\
   \vdots  & \ddots  & \vdots   \\
   {{c}_{N,0}} & \cdots  & {{c}_{N,N}}  \\
\end{matrix} \right)
\end{equation}
The initial placement is given in Eq. \ref{placement}. $\mathbb{P}_{s}$  is calculated using our previous optimization solution for VNF placement presented in \cite{globecomtamim}. For each $s$ (server) in $\mathbb{P}_{s}$, a list containing all the VNFs on that $s$ is provided. 
\begin{equation}
\label{placement}
{{\mathbb{P}}_{s}}=\{{{v}_{i{{d}_{0}}}},{{v}_{i{{d}_{1}}}},...{{v}_{i{{d}_{2}}}}\}
\end{equation}
In Equation \ref{eq:SFC}, each index of the set $\mathbb{S}\mathbb{F}\mathbb{C}$ contains a chain of three VNFs (near-RT RIC, O-CU, and O-DU), each VNF is a placed VNF from the set ${{\mathbb{P}}_{s}}$. Certain VNFs can exist in multiple SFC, and this is determined through the initial placement configuration provided by the NSP. The set $\mathbb{S}\mathbb{F}\mathbb{C}$ has at least $k=VNF count/3$ where 3 is the number of VNFs required to make an SFC in this use case, and no VNFs are shared. $k$ can become larger if certain VNFs are shared.
\begin{multline}
\label{eq:SFC}
\mathbb{S}\mathbb{F}\mathbb{C}=\{{{\{{{v}_{NR{{T}_{x}},}}{{v}_{OC{{U}_{y}},}}{{v}_{OD{{U}_{z}}}}\}}_{0}}, \\ 
{{\{{{v}_{NR{{T}_{x}},}}{{v}_{OC{{U}_{y}},}}{{v}_{OD{{U}_{z}}}}\}}_{1}},... \\ 
{{\{{{v}_{NR{{T}_{x}},}}{{v}_{OC{{U}_{y}},}}{{v}_{OD{{U}_{z}}}}\}}_{k}}\}\,\,,\,\, \\ 
where\,\,x,y,z\,\,\in \,\,{{\mathbb{P}}_{s}}
\end{multline}
The DRL agent achieves its objectives by estimating the near-optimal policy for the given O-RAN environment using $Q$-learning \cite{atari}. The near-optimal policy gives the agent the guideline in selecting a certain action at any state in the environment. The action space is defined as the set of those actions available to the agent. 

In our solution, we define the action space in Equation \ref{eq:ActionSpace} as the set of all possible TS decisions from a source VNF $v_{SRC}$ to a new destination VNF $v_{DST}$ where $SRC$ and $DST$ are VNF IDs that determine the VNF and the SFC it belongs to. Hence, the size of the action space is one plus $V$x$V$, where $V$ is the total number of VNFs in the O-RAN environment. The one additional action is the no TS action $NoTS$ where the agent does not issue any TS decision given that the network is operating within the acceptable queuing delay thresholds. 
\begin{multline}
\label{eq:ActionSpace}
ActionSpace=\{{{\{No\,TS\}}_{{{a}_{0}}}},{{\{{{v}_{SRC=0,}}{{v}_{DST=0}}\}}_{{{a}_{1}}}}, \\ {{\{{{v}_{SRC=0,}}{{v}_{DST=1}}\}}_{{{a}_{2}}}},...{{\{{{v}_{SRC=V,}}{{v}_{DST=V}}\}}_{{{a}_{(VxV)+1}}}}\}
\end{multline}

The state space the agent receives is defined in Equation \ref{eq:StateSpace}. The space consists of the sets $\mathbb{S}, \mathbb{L}, \mathbb{C},\mathbb{P}$ in addition to the current timestep $t$. The inclusion of the current timestep allows the agent to build knowledge about the status of the network and server infrastructure at a certain time of the day. This allows it to capture peak times ahead of their occurrence. For example, the agent builds knowledge that during the rush hours of the day, URLLC applications such as V2X are more prone to congestion and can incur higher queuing delays than the times of the night. This is the main contribution of the DRL agent; the ability to issue TS actions to optimal VNFs with respect to congestion and time. 
\begin{equation}
\label{eq:StateSpace}
State\,\,Space=\{ \mathbb{S}, \mathbb{L}, \mathbb{C},\mathbb{P},t\}
\end{equation}
\subsection{Reward Function and Deep Q-learning}
The reward function guides the agent's learning process by providing a scalar reward signal $r_t$ as feedback after the agent's every interaction with the environment. The goal of the DRL agent is to maximize the total reward $R$ after each episode $\tau$ of interaction with the environment, as shown in Eq. \ref{eq:TotalReward}. $\gamma$ is a discount factor in the range of [0, 1) \cite{discountfactor}.
\begin{equation}
\label{eq:TotalReward}
    R(\tau )=\sum\limits_{t=0}^{T}{\gamma {{r}_{t}}}
\end{equation}

The reward function is defined to cover four scenarios that the agent can face. At each timestep $t$, the agent chooses to issue a TS action or not from the action space. The policy the agent is aiming to reach is to only issue a TS action on the specific VNFs that are projected to face a queuing delay and to not issue a TS action when no congestion is expected. The agent is able to acquire such information from matrix $\mathbb{C}$.
\begin{figure*}[ht]
  \includegraphics[scale=0.5]{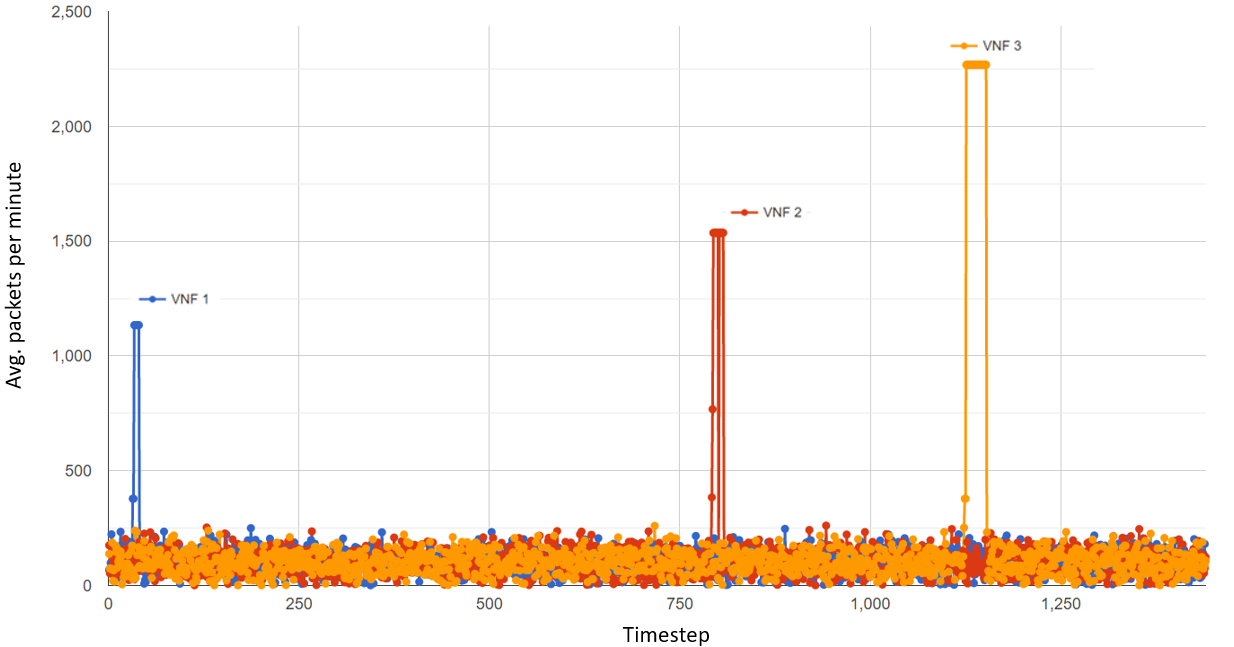}
  \setlength{\belowcaptionskip}{-10pt} 
  \caption{Spikes in packet arrival at three VNFs causing congestion}
    \label{fig:nbc}
\end{figure*}

To achieve this, we define the reward function in Equation \ref{eq:rewardfunction}. If action $a$ is $zero$ indicating no TS, the agent will receive a positive reward equal to one $\sigma$ if $c_{i,j}$ was zero $where\,\,c\in \mathbb{C},\,\,\,i,j\in {{\mathbb{P}}_{s}}$. $\sigma$ is a scalar value that determines the base of rewards the agents receive. We use $\sigma$ to clearly indicate the scale of how actions under specific scenarios are more favorable than others, and vice versa, actions under specific scenarios are more severe than others. The choice of a  no TS action when there exists a congestion results in a negative reward of eight times $\sigma$. $\sigma$ and the weights can be modified by the NSP to control the trade-off between the scenarios. 

When the agent chooses to issue a TS action, an overwhelmingly positive reward of ten times $\sigma$ is only issued when the source VNF is congested. On the other hand, if the agent issues a TS action on a non-congested VNF, it receives an overwhelmingly negative reward of fifteen times $\sigma$. 
\begin{equation}
\label{eq:rewardfunction}
{{r}_{t}}=
    \begin{cases} 
      \sigma  & \,\,\,if\,a=0\,,\sum\limits_{i=0}^{N}{\sum\limits_{j=0}^{N}{{{c}_{ij}}=0}}  \\
      -8\sigma  & \,\,\,if\,a=0,\,\sum\limits_{i=0}^{N}{\sum\limits_{j=0}^{N}{{{c}_{ij}}\ge 1}}  \\
      10\sigma  &\,\, if\,a={{a}_{v}}\,,{{c}_{\{{{v}_{src}}\}{{a}_{v}}}}=1\,,{{a}_{v}}>0  \\
      -15\sigma  &\,\, if\,a={{a}_{v}}\,,{{c}_{\{{{v}_{src}}\}{{a}_{v}}}}=0\,,{{a}_{v}}>0  \\
   \end{cases}
\end{equation}
The environment, agent, and rewarding mechanism are all part of the deep $Q$-learning approach. The goal of deep $Q$-learning is to alleviate the impracticality of representing the $Q$-function as a table containing values for each combination of states and actions. Instead, an artificial neural network is trained as a function approximator with parameters $\theta$. The $Q$-function then becomes $Q(s,a;\theta )\approx {{Q}^{*}}(s,a)$ \cite{atari} \cite{q}. Our deep neural network responsible for the agent consists of three hidden layers (24, 48, and 24 neurons per layer, respectively), and a learning rate of $0.005$. This configuration was selected through a grid search.  
\section{Simulation, Evaluation Settings, and Results}
\label{results}
\begin{figure}[t]
    \centering
    \setlength{\belowcaptionskip}{-20pt}
    \includegraphics[width=\columnwidth]{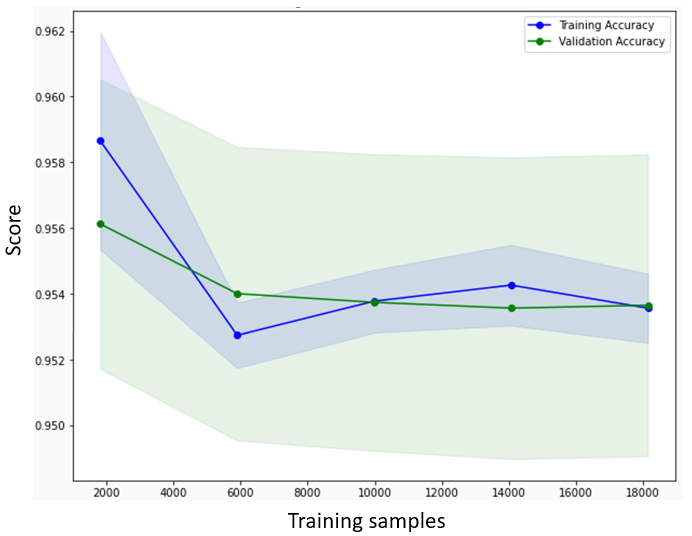}
    \caption{Training and validation accuracy of the NBC}
    \label{fig:nbctraining}
\end{figure}
\begin{figure}[htp]
    \centering
    \setlength{\belowcaptionskip}{-20pt}
    \includegraphics[width=\columnwidth]{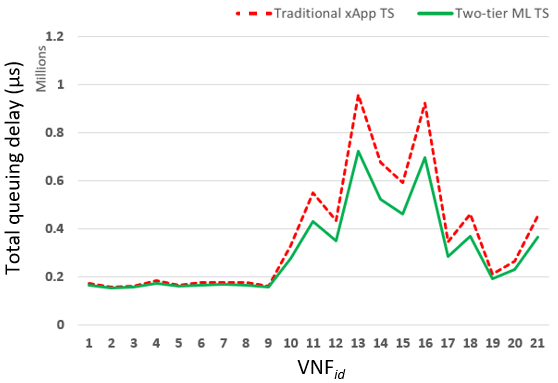}
    \caption{Daily queuing performance of traditional xApp TS versus our solution per VNF}
    \label{fig:tdelay}
\end{figure}
Our simulation infrastructures consist of 50 COTS and 21 VNFs (7 near-RT RICs, 7 O-CUs, and 7 O-DUs). To test both ML models of our solution, we simulate URLLC traffic (details mentioned in Section \ref{nbcsection}) requests as packets per second from UEs to every VNF for a 1 day period (1440 timesteps) through a discrete event simulator following the URLLC traffic specifications presented in \cite{TRAFFIC}. The generated traffic causes congestion at all VNFs throughout the day. Fig. \ref{fig:nbc} shows an example of three VNFs spiking congestion at one point of the day. The goal of the NBC is to be able to classify such congestion on a VNF given the average packets that are being processed in a minute for that VNF. 

Fig. \ref{fig:nbctraining} shows the training performance convergence of the NBC. To effectively capture the performance in terms of true positives and false positives, we calculate the AUC scores. Under testing, our NBC was able to achieve an AUC score of 0.978. This was sufficient to pass the generated congestion predictions to the agent. 

It is important to note that this score is high as our data is simulated. For a larger dataset acquired from a real-world O-RAN deployment, the model needs to be much further tuned to achieve similar high scores. We aim to address this challenge in our future work.  

The same simulation infrastructure is passed to the O-RAN RL environment. The environment ensures that no TS decision is accepted if it violates any of O-RAN operational and/or functional constraints. Our agent is an episodic deep $Q$-learning agent with memory reply \cite{REPLY}. One episode of the environment has 1440 steps corresponding to the minutes per day. The agent is able to guarantee convergence of the deep neural net after 14 episodes in the environment. Critically, and to determine the successful contribution of the agent, we observe that TS decisions are being issued before congestion occurs. Following the above-defined reward function, this causes the biggest positive reward. We observed this behaviour three out of the four times a significant congestion occurred during the testing.  
To compare the advantages our two-tier ML solution brings, Fig. \ref{fig:tdelay} shows the queuing delay incurred at each VNF over the day with the solution active versus traditional reactive TS xApp. Overall, our solution provides a 15.81 percent improvement in queuing delay, as shown. The training of the agent was completed in approximately $57h53m$ using an Intel 9th Gen I7-9750H 2.6 GHz CPU computing server with 16GB RAM. It is important to note that this training can be done offline and does not affect the operation of the O-RAN as the inference of the model can be done in less than $1s$ in the O-CU/O-DU.

For our future work, we aim to expand the RL environment to include further state parameters that will aid the agent in larger O-RAN deployments. In addition, we aim to deploy a real-world O-RAN and test our solution with a larger set of URLLC traffic. 

\section{Conclusion}
We have presented an ML-based TS mechanism for O-RAN capable of both detecting upcoming congestion-causing demands and rerouting them to the most appropriate SFC significantly reducing queuing delays and respecting O-RAN's operational and functional constraints. The system boasts high adaptability to changes in the network due to its $Q$-learning-based approach. It also allows for easier integration as it requires limited and locally available information collected by O-RANs DCAE. The performance of our proposed system has been evaluated using simulated URLLC datasets, and it has  shown to achieve consistently lower queuing delays (15.81 percent improvement in queuing delays) compared to traditional xApp TS-based methods for each O-RAN VNF.

\bibliographystyle{IEEEtran}
\bibliography{Biblio/bib}

\end{document}